# Phase-optimised linearly-constrained minimum-variance beamformers

Hugh L. Kennedy
STEM Unit
University of South Australia
Adelaide, Australia
hugh.kennedy@unisa.edu.au

*Abstract*—**A procedure for the determination of the optimal group-delay of a Linearly-Constrained Minimum-Variance (LCMV) beamformer is proposed. Two ways of selecting the optimal delay are recommended: the first is the solution that minimizes the noise power; the second is the solution that minimizes the processing delay. The potential of this hitherto unexplored degree of design freedom is investigated using simulated Very-High-Frequency (VHF) communication, and Ultra-High-Frequency (UHF) bistatic radar, applications.**

***Keywords—adaptive beamforming, optimal beamforming, multidimensional signal processing, spatiotemporal filtering, array processing, phased arrays***

## I. INTRODUCTION

The Linearly-Constrained Minimum-Variance (LCMV) beamformer is an adaptive processing technique for the suppression of interference in sensor arrays such as those found in modern communication and radar systems. Given the sampled covariance matrix of the Radio-Frequency (RF) environment, the beamformer adjusts its response to minimise the gain in the directions of interferers while maintaining unity gain in the look direction [1].

The covariance matrix may be estimated and inverted recursively online using the (constrained) Least Mean-Squares (LMS) algorithm [2]. This approach is appropriate when the RF environment is *variable* and the target direction is accurately known, e.g. in semi-active radar homing systems for single-target tracking. However, it is assumed here that RF interference is relatively *static* and that the target's direction is not precisely known, e.g. in a fixed-site bistatic radar system for surveillance and early warning, thus the covariance matrix is estimated (and curated) from long-term time-averages of the environment using regular survey dwells when targets are known to be absent. Covariance matrix estimation/approximation is a critical part of the adaptive array-processing problem; however, the primary focus of this paper is on the exploitation of the information contained therein, for the extraction of weak signals that would otherwise be undetectable in high-power correlated noise.

*Spatial* sampling, using a distributed array of omni-directional sensing elements, and *one-dimensional* filters are used to shape the beamformer's response so that gain is minimised at angles where interference power is greatest. The LCMV beamformer reduces to the Minimum-Variance Distortionless-Response (MVDR) beamformer in this limiting case. *Spatiotemporal* sampling, using a multi-channel analogue-to-digital converter (ADC) and *two-dimensional* filters are used to shape the response in both angle and time/frequency. These additional degrees of freedom may be used to correct the frequency dependence of the realized look direction (i.e. so-called "beam squint") for the high-gain reception of signals with a wide fractional bandwidth. They may also be used to supress narrowband interference that is in the look direction of a wideband signal but at a different centre frequency; however, this aspect of wideband LCMV beamforming is rarely discussed in the literature and investigations/simulations usually consider interferers and signals with the same centre frequency and similar bandwidth.

Early LCMV formulations assume that the array is (mechanically or electronically) pre-steered to place the desired signal at array broadside [2],[6]; however, simpler realisations are reached when the linear constraints incorporate complex coefficients to align signals arriving from arbitrary angles [3],[4],[7],[8],[9],[10]. In addition to the simple (unity) magnitude requirement at a given angle and frequency, further derivative constraints may be applied to shape the spatiotemporal response [1]. Specifying derivatives at critical operating frequencies, e.g. at $\omega = 0$ (dc) and $\omega = \pi$ (Nyquist), is a simple and effective technique for the design of low-order low-pass digital filters, with a Finite Impulse Response (FIR) or an Infinite Impulse Response (IIR) and an arbitrary group delay [12],[13],[14]. The resulting filters have a wide transition band for good high-frequency attenuation and a compact/well-damped impulse response. Wideband LCMV beamformers are usually designed by specifying the impulse [2], or frequency response at a series of discrete points [5],[7],[8],[9],[10]. Derivative constraints up to second order at many frequencies are used to design wideband beamformers in [11]. However, high-order derivative constraints at only dc and Nyquist are used here for the responses described above and because they lead to simple closed-form expressions that allow the group/phase delay of the digital filter to be optimised.

The *spatial* phase-centre of a uniform linear array with $M_s$ (omni-directional) elements is usually chosen to be at $q_s = (M_s - 1)/2$; however, this does not necessarily maximise the Signal to Noise Ratio (SNR). Expressions for the noise power as a function of $q_s$ may be derived when derivative constraints are imposed on the angle response in the look direction [6],[8]. The value of $q_s$ that minimises the noise power may then be chosen and used to design the spatial weights of the beamformer. An analogous approach in the temporal dimension does not appear to have been reported in the literature and a temporal phase centre of $q_t = (M_t - 1)/2$ is usually assumed, where $M_t$ is the number of tapped delays. The primary contribution of this paper is to present a simple procedure for the determination of the optimal *temporal* phase-centre for a wideband LCMV beamformer. Bulky and expensive analogue electronic components are required to improve angular resolution using an array with a greater aperture; however, in the age of very large-scale integration, improved frequency resolution is freely available using longer FIR filters, but long group delays may not be necessary. Thus the determination of optimal delays for LCMV beamformers is an area that deserves attention.

## II. Phase-Optimised LCMV Beamforming

The $n$th output of the beamformer $y[n]$ is found by convolving the sensor samples $x$ with the $M_t \times M_s$ beamformer weight matrix $\boldsymbol{H}$ using

$$y[n] = \sum_{m_s=0}^{M_s-1} \sum_{m_t=0}^{M_t-1} h[m_t, m_s] x[n - m_t, m_s] \tag{1}$$

where

$M_s$ is the number of antennas,

$M_t$ is the number of delays and

$h[m_t, m_s]$ are the elements of $\boldsymbol{H}$.

The optimal weight-matrix coefficients of the LCMV beamformer are determined by minimising the noise power

$$P = \boldsymbol{h}^\dagger \boldsymbol{R} \boldsymbol{h} \tag{2a}$$

subject to the linear constraints

$$\boldsymbol{d} = \boldsymbol{F}^\dagger \boldsymbol{h} \tag{2b}$$

where

$\boldsymbol{R}$ is the $M_s M_t \times M_s M_t$ spatiotemporal noise covariance matrix,

$\boldsymbol{h}$ is the weight vector with the columns of $\boldsymbol{H}$ stacked as

$$\boldsymbol{h} = \begin{bmatrix} h[0,0] \\ \vdots \\ h[M_t - 1, 0] \\ \vdots \\ h[0, M_s - 1] \\ \vdots \\ h[M_t - 1, M_s - 1] \end{bmatrix}_{M_s M_t \times 1} \tag{2c}$$

and

$\blacksquare^\dagger$ is the Hermitian transpose operator.

The constraints are defined using the $K \times 1$ vector $\boldsymbol{d}$ and the $M_s M_t \times K$ matrix $\boldsymbol{F}$, where $K$ is the total number of constraints. The beamformer's response $\mathcal{H}(\omega_t, \theta)$ is a function of *temporal* frequency $\omega_t$ (radians per temporal sample) and angle $\theta$ (radians, where $\theta = 0$ is perpendicular to the linear array, i.e. 'broadside'). It is assumed that the signal has been mixed down to baseband so that its spectrum is centred on $\omega_t = 0$ (thus $y$, $h$, $x$, and the elements of $\boldsymbol{R}$, are complex). Its bandwidth and passband group-delay may therefore be specified using only $k$th partial derivatives with respect to frequency $\partial^k \mathcal{H}(\omega_t, \theta_l)/\partial \omega_t^k$, in the look direction $\theta_l$, for $k = 0 \ldots K - 1$, with $K = K_{dc} + K_{pi}$, where $K_{dc}$ and $K_{pi}$ are the number of derivative constraints at $\omega_t = 0$ and $\omega_t = \pi$, respectively. Wideband responses are obtained using large $K_{dc}$ and small $K_{pi}$. Partial derivatives with respect to angle are not considered here, other than $k = 0$ for unity gain in the look direction. This allows the beam to distort, with a left or right skew, to avoid wideband interferers that are at the signal's carrier frequency and close to the look direction. For a low-frequency group-delay of $q_t$ samples, the $K$ constraints on the beamformer's frequency response are

$$\boldsymbol{d} = \begin{bmatrix} \boldsymbol{d}_{dc} \\ \boldsymbol{d}_{pi} \end{bmatrix}_{K \times 1} \tag{3a}$$

where

the $k$th element of $\boldsymbol{d}_{dc}$ is

$$\left.\frac{\partial^k \mathcal{H}(\omega_t, \theta_l)}{\partial \omega_t^k}\right|_{\omega_t = 0} = \left(-i q_t\right)^k \quad \text{for } 0 \le k < K_{dc} \text{ and} \tag{3b}$$

the $k$th element of $\boldsymbol{d}_{pi}$ is

$$\left.\frac{\partial^k \mathcal{H}(\omega_t, \theta_l)}{\partial \omega_t^k}\right|_{\omega_t = \pi} = 0 \qquad \text{for } 0 \le k < K_{pi} \tag{3c}$$

The beamformer's two-dimensional impulse response is equal to the weight matrix $\boldsymbol{H}$, thus it's response in temporal frequency and angle is evaluated using

$$\mathcal{H}(\omega_t, \theta) = \sum_{m_s=0}^{M_s-1} \sum_{m_t=0}^{M_t-1} h[m_t, m_s] e^{i(m_s \omega_s \sin\theta - m_t \omega_t)} \tag{4a}$$

where

$i$ is the imaginary unit

$\omega_s$ is the spatial frequency (radians per spatial sample) which is a constant, with

$$\omega_s = 2\pi F_c D / v_c \tag{4b}$$

where

$F_c$ is the centre frequency of the signal, i.e. the radio-frequency carrier (cycles per second, i.e. Hz),

$D$ is the distance between adjacent elements of the uniform linear array (metres per spatial sample) and

$v_c$ is the velocity of the signal carrier, i.e. the speed of light (metres per second).

For an unambiguous angular response, the array should be designed with $D \le \lambda_c/2$, where $\lambda_c$ is the wavelength (meters per cycle) of the signal carrier.

To evaluate the elements of the $M_s M_t \times K$ matrix $\boldsymbol{F}$ in (2b), let the element in the $m_t$th row and the $m_s$th column of the $M_t \times M_s$ matrix $\boldsymbol{\mathcal{F}}$, be the basis function $f(\omega_t, \theta; m_t, m_s) = e^{i(m_t \omega_t - m_s \omega_s \sin\theta)}$, and let $\boldsymbol{f}$ be a vector formed by stacking the columns of $\boldsymbol{\mathcal{F}}$ such that

$$\boldsymbol{f} = \begin{bmatrix} f(\omega_t, \theta; 0,0) \\ \vdots \\ f(\omega_t, \theta; M_t - 1,0) \\ \vdots \\ f(\omega_t, \theta; 0, M_s - 1) \\ \vdots \\ f(\omega_t, \theta; M_t - 1, M_s - 1) \end{bmatrix}_{M_s M_t \times 1} . \tag{5}$$

The matrix $\boldsymbol{F}$ is then constructed using

$\boldsymbol{F} = [\boldsymbol{F}_{dc} \quad \boldsymbol{F}_{pi}]$ with

$$\boldsymbol{F}_{dc} = \begin{bmatrix} \left.\frac{\partial^0 f(0,0)}{\partial \omega_t^0}\right|_{\omega_t=0} & \cdots & \left.\frac{\partial^k f(0,0)}{\partial \omega_t^k}\right|_{\omega_t=0} & \cdots & \left.\frac{\partial^{K_{dc}-1} f(0,0)}{\partial \omega_t^{K_{dc}-1}}\right|_{\omega_t=0} \\ \vdots & & \vdots & & \vdots \\ \left.\frac{\partial^0 f(M_t-1,0)}{\partial \omega_t^0}\right|_{\omega_t=0} & \cdots & \left.\frac{\partial^k f(M_t-1,0)}{\partial \omega_t^k}\right|_{\omega_t=0} & \cdots & \left.\frac{\partial^{K_{dc}-1} f(M_t-1,0)}{\partial \omega_t^{K_{dc}-1}}\right|_{\omega_t=0} \\ \vdots & & \vdots & & \vdots \\ \left.\frac{\partial^0 f(0,M_s-1)}{\partial \omega_t^0}\right|_{\omega_t=0} & \cdots & \left.\frac{\partial^k f(0,M_s-1)}{\partial \omega_t^k}\right|_{\omega_t=0} & \cdots & \left.\frac{\partial^{K_{dc}-1} f(0,M_s-1)}{\partial \omega_t^{K_{dc}-1}}\right|_{\omega_t=0} \\ \vdots & & \vdots & & \vdots \\ \left.\frac{\partial^0 f(M_t-1,M_s-1)}{\partial \omega_t^0}\right|_{\omega_t=0} & \cdots & \left.\frac{\partial^k f(M_t-1,M_s-1)}{\partial \omega_t^k}\right|_{\omega_t=0} & \cdots & \left.\frac{\partial^{K_{dc}-1} f(M_t-1,M_s-1)}{\partial \omega_t^{K_{dc}-1}}\right|_{\omega_t=0} \end{bmatrix}_{M_s M_t \times K_{dc}} \tag{6a}$$

and

$$\boldsymbol{F}_{pi} = \begin{bmatrix} \left.\frac{\partial^0 f(0,0)}{\partial \omega_t^0}\right|_{\omega_t=\pi} & \cdots & \left.\frac{\partial^k f(0,0)}{\partial \omega_t^k}\right|_{\omega_t=\pi} & \cdots & \left.\frac{\partial^{K_{pi}-1} f(0,0)}{\partial \omega_t^{K_{pi}-1}}\right|_{\omega_t=\pi} \\ \vdots & & \vdots & & \vdots \\ \left.\frac{\partial^0 f(M_t-1,0)}{\partial \omega_t^0}\right|_{\omega_t=\pi} & \cdots & \left.\frac{\partial^k f(M_t-1,0)}{\partial \omega_t^k}\right|_{\omega_t=\pi} & \cdots & \left.\frac{\partial^{K_{pi}-1} f(M_t-1,0)}{\partial \omega_t^{K_{pi}-1}}\right|_{\omega_t=\pi} \\ \vdots & & \vdots & & \vdots \\ \left.\frac{\partial^0 f(0,M_s-1)}{\partial \omega_t^0}\right|_{\omega_t=\pi} & \cdots & \left.\frac{\partial^k f(0,M_s-1)}{\partial \omega_t^k}\right|_{\omega_t=\pi} & \cdots & \left.\frac{\partial^{K_{pi}-1} f(0,M_s-1)}{\partial \omega_t^{K_{pi}-1}}\right|_{\omega_t=\pi} \\ \vdots & & \vdots & & \vdots \\ \left.\frac{\partial^0 f(M_t-1,M_s-1)}{\partial \omega_t^0}\right|_{\omega_t=\pi} & \cdots & \left.\frac{\partial^k f(M_t-1,M_s-1)}{\partial \omega_t^k}\right|_{\omega_t=\pi} & \cdots & \left.\frac{\partial^{K_{pi}-1} f(M_t-1,M_s-1)}{\partial \omega_t^{K_{pi}-1}}\right|_{\omega_t=\pi} \end{bmatrix}_{M_s M_t \times K_{pi}} \tag{6b}$$

where the continuous $\omega_t$ and $\theta$ arguments of the basis functions have been dropped for brevity, leaving only the integer parameters $m_t$ and $m_s$. These partial derivatives are readily evaluated using

$$\left.\frac{\partial^k f(m_t,m_s)}{\partial \omega_t^k}\right|_{\omega_t=0} = (im_t)^k e^{-im_s\omega_s \sin\theta_l} \text{ and} \tag{7a}$$

$$\left.\frac{\partial^k f(m_t,m_s)}{\partial \omega_t^k}\right|_{\omega_t=\pi} = (im_t)^k e^{-im_s\omega_s \sin\theta_l}(-1)^{m_t} . \tag{7b}$$

Following the approach used in [12] to design one-dimensional IIR smoothers: for a specified group-delay $q_t$ the optimal weights in (2c) are obtained by minimising (2a) subject to (2b) using

$$\boldsymbol{h} = \boldsymbol{R}^{-1}\boldsymbol{F}(\boldsymbol{F}^{\dagger}\boldsymbol{R}^{-1}\boldsymbol{F})^{-1}\boldsymbol{d} . \tag{8}$$

For an unspecified group-delay, (3b) indicates that the noise power in (2a) is a function of $q_t$, i.e.

$$P(q_t) = \boldsymbol{h}(q_t)^{\dagger}\boldsymbol{R}\boldsymbol{h}(q_t) . \tag{9}$$

The following equivalent expression may also be used

$$P(q_t) = \boldsymbol{d}^\dagger(q_t)(\boldsymbol{F}^\dagger\boldsymbol{R}^{-1}\boldsymbol{F})^{-1}\boldsymbol{d}(q_t) \tag{10}$$

which allows $P(q_t)$ to be analysed, and the optimal value of the group delay selected, without first solving (2). An obvious way to select $q_t$ is to minimise $P(q_t)$ by solving

$$P'(q_t) = 0 \tag{11a}$$

where

$$P'(q_t) = dP(q_t)/dq_t. \tag{11b}$$

Inspection of (3b) indicates that (10) is simply solved by finding the roots of a polynomial with a degree of $2(K_{dc} - 1) - 1$. The roots should ideally all be positive real values; however, ill conditioned matrices yield non-negligible imaginary parts or negative real values, and these solutions should be ignored. In the absence of a feasible solution $q_t = (M_t - 1)/2$ should be used as a fallback. An alternative approach to the selection of the optimal group delay is to minimise the latency of the beamformer by choosing the feasible root with the smallest (positive) real value. Both approaches are investigated in the simulations that follow.

## III. SIMULATIONS

Sequences of phase-coded Slepian-like pulses were used to analyse various beamformers in communication and bistatic radar applications, in the very-high frequency (VHF) and ultra-high frequency (UHF) bands, respectively. Slepian tapers allow the trade-off between sidelobe-height and main-lobe width to be controlled by maximising energy concentration within the specified low-frequency band; thus Slepian pulse-shaping filters reduce out-of-band interference and lower bit-error rates in communication systems [15]. For covert radar operation, this property also lowers the probability of intercept and detection in the frequency domain. Using quasi-random communication-like waveforms in active bistatic radar systems also increases the probability of misclassification (as a communication signal) if detected. Moreover, they have a lower probability of detection in the time domain, relative to conventional pulsed waveforms of equivalent average power, because signal energy is spread more uniformly over time, thus peak energy is low. Furthermore, their so-called "thumb-tack" ambiguity functions in delay and Doppler have a sharp main-lobe and an absence of range/velocity ambiguities. However, their flat side-lobe structure, which does not roll off with increasing delay, increases clutter susceptibility; regardless, digital communication waveforms are ideal for passive bistatic radars.

The pulse-shaping filters used here, to design (communication and radar) waveforms with a bandwidth of $B_{pls}$ (cycles per second, i.e. Hz) were designed by minimising the stopband power i.e. for $2\pi B_{pls}/F_{smp} \le |\omega| \le \pi$, where $F_{smp}$ is the sampling rate (samples per second), subject to $K_{pls}$ derivative constraints at $\omega = 0$. For these pulses, $B_{pls}$ is the desired first-null bandwidth, not the more conventional 3 dB bandwidth. For a given pulse-repetition interval of $M_{pls}$ samples or $T_{pls} = M_{pls}/F_{smp}$ seconds, using $K_{pls} \gg 1$, yields a wider 3 dB bandwidth but increases the out-of-band power. A single derivative constraint was used ($K_{pls} = 1$) for Slepian-like pulses. The baseband signal is then used to modulate a carrier of $F_{pls}$ Hz.

The pulse-shaping filter is convolved with a sequence of complex impulses $e^{i\phi}$ where $\phi$ is the pulse phase (radians) which was randomly assigned to one of four possibilities ($0$, $\pi/4$, $\pi/2$ or $3\pi/4$) with equal probability. Each impulse is followed by a sequence of $M_{pls} - 1$ zeros.

In addition to the signal transmitter, two other types of emitters were simulated: interferers and jammers. Interferers are narrowband emitters that are close to the steering direction and within the signal's frequency band; whereas jammers are wideband high-power emitters that are on average further from the steering direction with the same centre frequency as the signal. For these emitters, the angle ($\theta$), average magnitude ($A$), first-null bandwidth ($B$), and centre-frequency ($F$) parameters are denoted using the $\blacksquare_{int}$ and $\blacksquare_{jam}$ subscripts, respectively. Their waveforms were generated by passing complex zero-mean Gaussian noise through a low-pass noise-shaping filter. The procedure used to design the signal pulse-shaping filter was also used to design the interferer and jammer noise-shaping filters. They have a length of $M_{int}$ and $M_{jam}$ samples, with $K_{int}$ and $K_{jam}$ derivative constraints at $\omega = 0$, and a single derivative constraint at $\omega = \pi$. These baseband waveforms are used to modulate the carriers at $F_{int}$ and $F_{jam}$ Hz, where $F_{int}$ was drawn randomly from a uniform distribution over the $\left(F_{int}^{min}, F_{int}^{max}\right)$ interval, where $F_{int}^{min} = F_{pls} - B_{pls}/2$ and $F_{int}^{max} = F_{pls} + B_{pls}/2$; whereas $F_{jam} = F_{pls}$. The angles of the interferers and jammers were randomly generated using a uniform distribution over the $\left(\theta_{int}^{min}, \theta_{int}^{max}\right)$ and $\left(\theta_{jam}^{min}, \theta_{jam}^{max}\right)$ intervals, respectively. In each random instantiation of the communication and radar scenarios, $N_{int}$ interferers and $N_{jam}$ jammers were randomly generated (see TABLE I. ).

The waveforms at each antenna element were simulated by applying an appropriate phase-shift to the carrier and by using the dc-derivative constraints in (3b) to delay the baseband signal. Random zero-mean Gaussian-distributed (white) noise was added to each yield an input SNR of -30 dB (excluding interferer and jammer contributions). The antenna elements only receive the real part of the radio-frequency waveform. In an actual system, an analogue band-pass filter would be applied to extract only the RF band of interest, centred on $F_{pls}$. However, in these simulations there is nothing outside the band of interest to attenuate. The waveform is sampled by the ADC at a rate of $F_{smp}$ Hz. As this rate is less than all RF frequencies, the waveform is aliased into the bandwidth of the ADC, i.e. over the $\pm F_{smp}/2$ interval. The real discrete-time waveform is mixed with an complex oscillator at -10 MHz to shift it down to 0 Hz and a low-pass FIR filter is applied to the now complex discrete-time waveform, yielding $x$ in (1). The FIR filter is relatively flat over the signal's passband. It has a length of 13 samples and is designed using 8 derivative constraints at $\omega = 0$, 5 derivative constraints at $\omega = \pi$, and a first-null bandwidth of $B_{pls}$.

In both communication and radar scenarios, $N_{smp}$ samples were collected for a dwell duration of $N_{smp}/F_{smp}$ seconds (see TABLE II. ). The look direction $(\theta_l)$ used for the beamformer was randomly generated using a uniform distribution over the $\left(\theta_{rxr}^{min}, \theta_{rxr}^{max}\right)$ interval. Signals of interest (i.e. the transmitter in the communication scenario and target returns in the radar scenario) were offset from $\theta_l$ using randomly generated angular errors that were normally distributed with a mean of zero and a standard deviation of $\sigma_\theta$. The six beamforming algorithms (BMF-A through to BMF-F) used to process the data are described in the subsections that follow. In both communication and radar scenarios, a survey dwell (with signals absent) was used to collect $N_{smp}$ samples for the establishment of the spatiotemporal noise covariance matrix ($\boldsymbol{R}$). The average SNR achieved for six beamformers was evaluated using 1000 and 100 random instantiations of the VHF communication and UHF radar scenarios, respectively.

TABLE I. SIMULATION PARAMETERS FOR THE VHF COMMUNICATION AND UHF RADAR SCENARIOS

| ***Parameter*** | ***Value*** |
|---|---|
| | |
| $F_{pls}$ | *250 MHz or 2.41 GHz* |
| $B_{pls}$ | *20 MHz* |
| $M_{pls}$ | *13* |
| $K_{pls}$ | *1* |
| | |
| $N_{int}$ | *4* |
| $F_{int}^{min}$ | *250 MHz - 20 MHz or 2.41 GHz - 20 MHz* |
| $F_{int}^{max}$ | *250 MHz + 20 MHz or 2.41 GHz + 20 MHz* |
| $B_{int}$ | *1 MHz* |
| $A_{int}$ | *100 mV* |
| $M_{int}$ | *240* |
| $K_{int}$ | *5* |
| $\theta_{int}^{min}$ | $\theta_l$ *- 30°* |
| $\theta_{int}^{max}$ | $\theta_l$ *+ 30°* |
| | |
| $N_{jam}$ | *1* |
| $F_{jam}$ | *250 MHz or 2.41 GHz* |
| $B_{jam}$ | *40 MHz* |
| $A_{jam}$ | *1000 mV* |
| $M_{jam}$ | *13* |
| $K_{jam}$ | *5* |
| $\theta_{jam}^{min}$ | *45°* |
| $\theta_{jam}^{max}$ | *90°* |
| | |
| $\sigma_{\theta}$ | *2°* |
| $\theta_{rxr}^{min}$ | *- 60°* |
| $\theta_{rxr}^{max}$ | *30°* |

TABLE II. PROCESSING PARAMETERS FOR THE VHF COMMUNICATION AND UHF RADAR SCENARIOS

| ***Parameter*** | ***Value*** |
|---|---|
| $F_{smp}$ | *40 MHz* |
| $N_{smp}$ | *100,000* |
| $M_s$ | *8* |
| *D* | *0.6 m or 0.0622 m* |

### A. *MVDR Beamformer*

Beamformer A (BMF-A) has no temporal degrees of freedom ($M_t = 1$) and is only constrained to have unity gain in the look direction.

### B. *Wideband MVDR Beamformer*

Beamformer B (BMF-B) applies MVDR processing to $M_f$ independent frequency channels, centred on $\omega_t = 0$. The channels are extracted using an $M_{DFT}$-point Discrete Fourier Transform (DFT) with $M_{DFT} \geq M_f$. The Fast Fourier Transform (FFT) is usually used for this purpose; however, complexity is reduced at the expense of increased latency due to its batch structure. A Kaiser window function, with a null-to-null main-lobe width that is twice as wide as the rectangular window function, is applied during the DFT to smooth beamformer's response, for a frequency resolution of $2F_{smp}/M_{DFT}$. The waveform in each channel is synthesized by evaluating each DFT bin at $m_t = M_{DFT} - K_f$. The bank of linear-phase FIR filters used for this analysis/synthesis operation may be applied recursively using the sliding DFT; however, non-recursive realisations have similar complexity for the short DFTs used here. The simulated scenarios were processed using $M_{DFT} = 9$, to approximately match the complexity of the LCMV beamformers; and $M_f = 7$, to approximately match the bandwidth of the signal.

### C. *Wideband LCMV Beamformer*

Beamformer C (BMF-C) applies LCMV processing to minimise the spatiotemporal noise power ($P$) using ($M_t$) temporal and ($M_s$) spatial degrees of freedom, subject to $M_f$ unity-gain constraints, in the look direction ($\theta_l$), at the centre of each frequency bin. Complex constraints are used to eliminate the need for pre-steering filters. Using $M_t = 9$ and $M_f = 5$ was found to be ideal for the scenarios considered here.

### D. *LCMV Beamformer with Derivative Constraints*

Beamformer D (BMF-D) applies LCMV processing, using ($K = K_{dc} + K_{pi}$) frequency-derivative constraints, as specified in (3)-(8) above. Using $M_t = 9$, $K_{dc} = 5$ and $K_{pi} = 2$ was found to be ideal.

### E. *LCMV Beamformer with Derivative Constraints and Noise-Power Minimisation*

Beamformer E (BMF-E) is the same as BMF-D; however, the group delay that minimises the noise power is used to design the filter coefficients ($\boldsymbol{h}$), as specified using (11) above.

### F. *LCMV Beamformer with Derivative Constraints and Group-Delay Minimisation*

Beamformer F (BMF-F) is the same as BMF-D; however, the minimum (feasible) group delay is used to design the filter coefficients ($\boldsymbol{h}$).

### G. *VHF Communication Scenario*

In these simulations a signal carrier ($F_{pls}$) of 250 MHz was used. The signal transmitter was approximately aligned (offset using Gaussian angular errors $\sigma_\theta$) with the look direction of the beamformer. In this scenario, the signal reaches the ADC with magnitude of $A_{pls}$ =1 mV. The received signal was not demodulated and bit error rates were not computed; instead, the signal was treated as a single symbol from a very large constellation. The transmitted signal $s[n]$ was least-squares fitted to the beamformed output $y[n]$. The power of the signal estimate $P_{est}$ is the sum of the squared fit $|\hat{s}[n]|^2$; whereas the error power $P_{err}$ is sum of squared differences $\varepsilon^2[n] = |\hat{s}[n] - s[n]|^2$. The achieved SNR, averaged over all random scenario instantiations, of the beamformers in the communication scenario was then evaluated using the ratio of $P_{est}$ to $P_{err}$ (see TABLE III. ).

TABLE III. AVERAGE SNR AND GROUP DELAYS FOR VARIOUS BEAMFORMERES IN THE VHF COMMUNICATION SCENARIO

| *Beamformer* | *SNR (dB)* | $q_t$ |
|---|---|---|
| *BMF-A* | *24.6070* | *0.0000* |
| *BMF-B* | *25.4569* | *4.0000* |
| *BMF-C* | *29.9041* | *4.0000* |
| *BMF-D* | *31.3876* | *4.0000* |
| *BMF-E* | *32.2655* | *3.8316* |
| *BMF-F* | *27.6065* | *0.7549* |

### H. *UHF Radar Scenario*

In these simulations a signal carrier ($F_{pls}$) of 2.41 GHz was used. An active bi-static radar system was loosely modelled. The signal transmitter was assumed to be slaved to the receiver so the (reference) signal $s[n]$ is known precisely; however, the location of the transmitter is not known. The transmitted signal reaches the ADC with a magnitude of $A_{txr}$, and as the true signal is already available, it is considered to be an additional source of noise; thus it should be suppressed by the beamformer and its received emissions included in $\boldsymbol{R}$. The angle of the transmitter $\theta_{txr}$ was randomly generated using a uniform distribution over the $\left(\theta_{txr}^{min}, \theta_{txr}^{max}\right)$ interval. The separation between the receiver and transmitter is assumed to be large enough for reasonable direct-path attenuation but close enough for a monostatic geometry to be assumed when measuring range and range rate.

In addition to the noise generated by the three types of active emitters (i.e. the transmitter, interferers and jammers), returns generated by three types of scatterers (of the transmitter signal) were added to the waveform: targets, clutter and reflectors. Parameters relating to the scatterers are denoted using the $\blacksquare_{tgt}$, $\blacksquare_{clt}$ and $\blacksquare_{rfl}$ subscripts, respectively. All scatterers are assumed to be close to the look direction. Targets are assumed to be airborne with ranges and radial velocities (i.e. range-rates) distributed randomly and uniformly over $\left(r_{tgt}^{min}, r_{tgt}^{max}\right)$ and $\left(v_{tgt}^{min}, v_{tgt}^{max}\right)$ intervals, respectively; and with angles distributed randomly and normally with a mean of $\theta_l$ and a standard derivation of $\sigma_\theta$. Clutter (numerous, small and far) and reflectors (few, large and close) are assumed to be stationary ($v_{clt} = v_{rfl} = 0$). Their ranges are distributed randomly and uniformly over the $\left(r_{clt}^{min}, r_{clt}^{max}\right)$ and $\left(r_{rfl}^{min}, r_{rfl}^{max}\right)$ intervals, respectively. In this radar scenario, the objective is to maximise the SNR of the target returns while minimising the impact of the clutter and reflectors, in addition to the suppressing the emitters described in Sub-Section G. Scatterers were excluded from the survey dwell that was used to estimate $\boldsymbol{R}$. In this scenario, the target, clutter, and reflector, returns reach the ADC with magnitudes of $A_{tgt}$, $A_{clt}$, and $A_{rfl}$, respectively; $N_{tgt}$, $N_{clt}$, and $N_{rfl}$ scatterers were randomly generated and included in each random instantiation of the scenario (TABLE IV. ).

For the collected dwell, range-Doppler processing was performed as follows: 1) Down-sample the output of the beamformer by a factor of two; 2) Suppress scatterers at zero Doppler (to prevent their sidelobes from hiding targets) using the reference waveform and a Wiener filter; 3) Evaluate range-velocity cells using time-delayed and Doppler shifted reference waveforms in a matched filter-bank (see TABLE V. ).

For the radar scenario, the achieved SNR for the various beamformers was evaluated (see TABLE VI. ) by dividing the average signal power by the average noise power. These quantities were computed using the squared magnitude for all cells that contain a target and don't contain a target, respectively. Zero-Doppler cells and cells surrounding each target cell were excluded from the analysis.

TABLE IV. SIMULATION PARAMETERS FOR THE UHF RADAR SCENARIO

| ***Parameter*** | ***Value*** |
|---|---|
| | |
| $A_{txr}$ | *1,000 mV* |
| $\theta_{txr}^{min}$ | *45º* |
| $\theta_{txr}^{max}$ | *90º* |
| | |
| $N_{tgt}$ | *16* |
| $A_{tgt}$ | *1 mV* |
| $r_{tgt}^{min}$ | *1,000 m* |
| $r_{tgt}^{max}$ | *10,000 m* |
| $v_{tgt}^{min}$ | *-500 m/s* |
| $v_{tgt}^{max}$ | *-50 m/s* |
| | |
| $N_{clt}$ | *24* |
| $A_{clt}$ | *100 mV* |
| $r_{clt}^{min}$ | *1,000 m* |
| $r_{clt}^{max}$ | *10,000 m* |
| $v_{clt}$ | *0 m/s* |
| | |
| $N_{rfl}$ | *4* |
| $A_{rfl}$ | *1,000 mV* |
| $r_{rfl}^{min}$ | *100 m* |
| $r_{rfl}^{max}$ | *1,000 m* |
| $v_{rfl}$ | *0 m/s* |

TABLE V. RANGE-DOPPLER PROCESSING PARAMETERS FOR THE UHF RADAR SCENARIO

| ***Parameter*** | ***Value*** |
|---|---|
| *Range Resolution* | *7.5 m* |
| *Number of Range Cells* | *1,337* |
| *Minimum Range* | *0 m* |
| *Maximum Range* | *10,020 m* |
| *Velocity Resolution* | *24.8963 m/s* |
| *Number of Velocity Cells* | *49* |
| *Minimum/Maximum Velocity* | *±597.5104 m/s* |

TABLE VI. AVERAGE SNR AND GROUP DELAYS FOR VARIOUS BEAMFORMERES IN THE UHF RADAR SCENARIO

| ***Beamformer*** | ***SNR (dB)*** | $q_t$ |
|---|---|---|
| *BMF-A* | *13.3184* | *0.0000* |
| *BMF-B* | *15.4497* | *4.0000* |
| *BMF-C* | *16.1123* | *4.0000* |
| *BMF-D* | *16.3562* | *4.0000* |
| *BMF-E* | *16.6842* | *3.8824* |
| *BMF-F* | *15.8219* | *0.7519* |

## IV. DISCUSSION

In both the communication and radar scenarios, the SNR increases from BMF-A through to BMF-E (see TABLE III. & TABLE VI. ), and there is an SNR penalty for the reduced delay of BMF-F (less than one sample); however, its SNR is more than 2 dB greater than BMF-A (with no delay). The different constraints imposed by the various beamformers force their responses to ‘fold’ in different ways to mitigate environmental noise by placing interferers and jammers in (or near) magnitude ‘troughs’. The absence of temporal degrees of freedom in BMF-A means that its response can only fold in the angle dimension which increases interferer gain in these scenarios (see Figure 1). The temporal degrees of freedom in BMF-B through to BMF-F allow their responses to fold more freely in both angle *and* frequency (see Figure 2, Figure 3 & Figure 4). The higher average SNR of BMF-D relative to BMF-C (more pronounced in the communication scenario) is probably due to the use of 'soft' derivative constraints at frequency extremes rather than 'hard' constraints at discrete frequencies within the signal band, which may inhibit the formation of troughs in the look direction near the selected bins. The average SNR of BMF-B in the communication scenario increases more than 3 dB (to 29.0377) when the length of the DFT is increased, using $M_{DFT} = 15$ and $M_f = 11$ (for approximately the same bandwidth). The average SNR of BMF-C decreases by more than 3 dB (to 26.5565) when $M_t - M_f$ zero-magnitude constraints are applied in the stopband in addition to the $M_f$ unity-magnitude constraints in the passband.

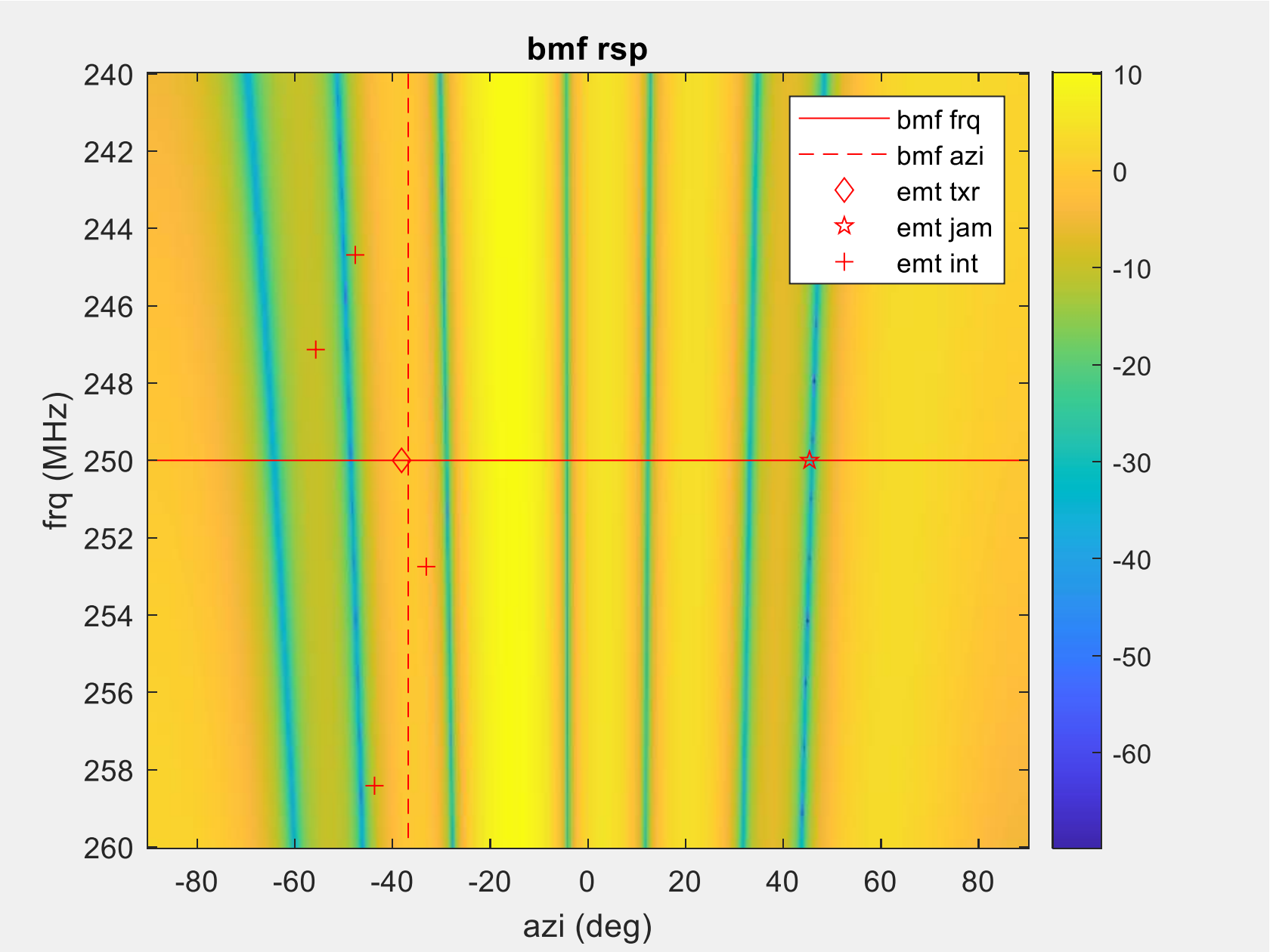

*Figure 1. Magnitude response (dB) of BMF-A; SNR = 4.9789 dB in this random instantiation.*

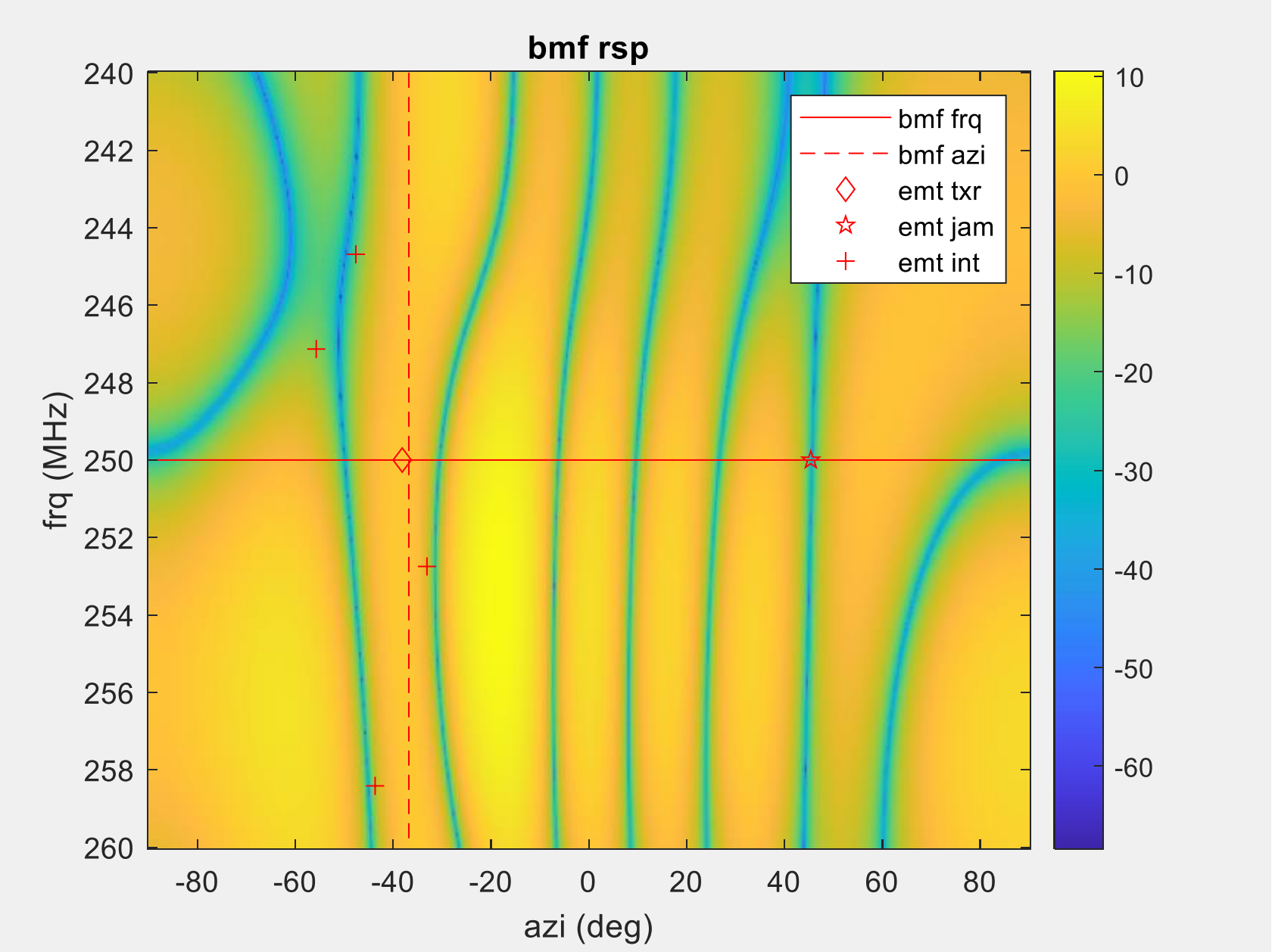

*Figure 2. Magnitude response (dB) of BMF-B; SNR = 9.5159 dB in this random instantiation.*

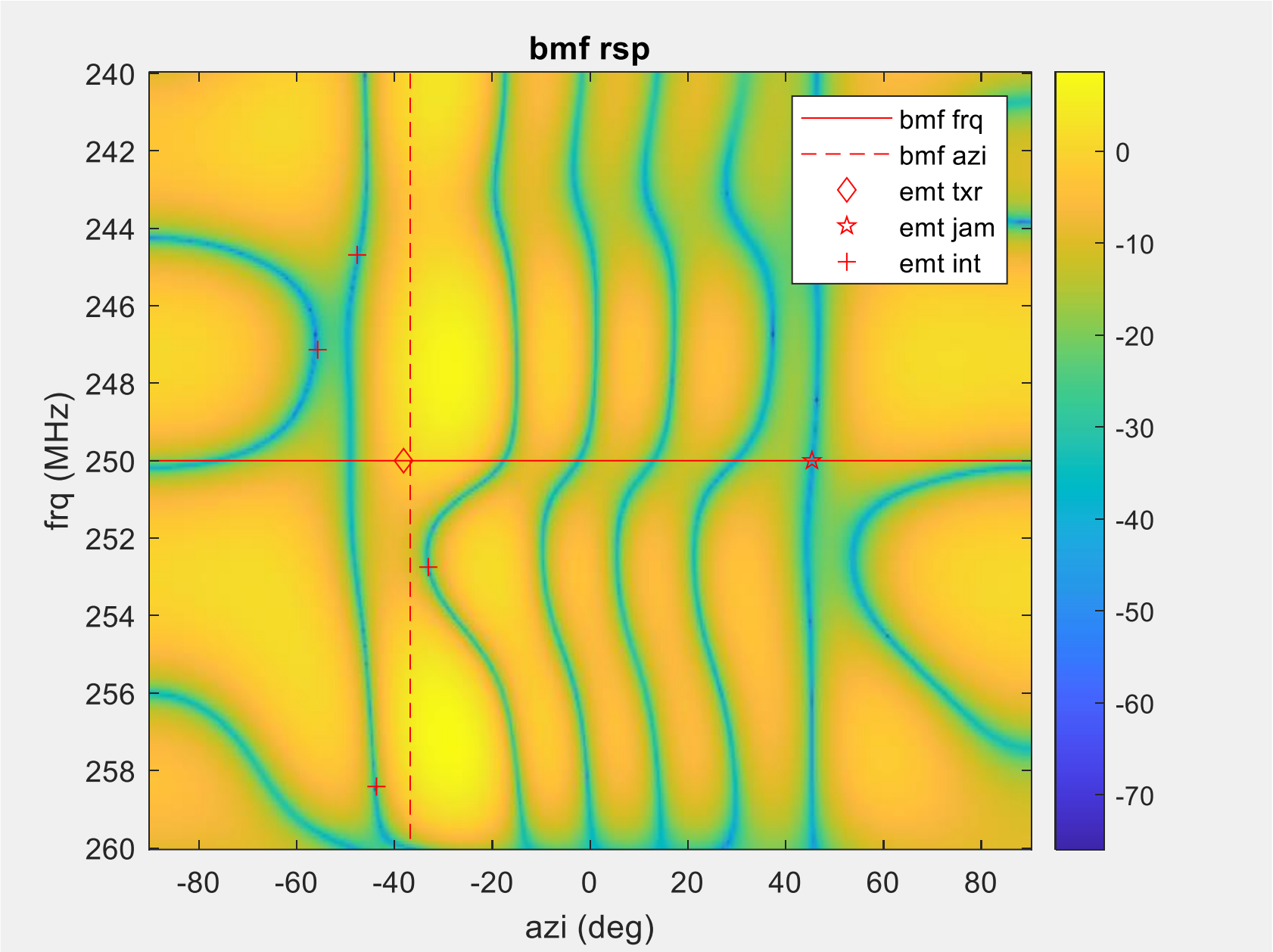

*Figure 3. Magnitude response (dB) of BMF-C; SNR = 15.0947 dB in this random instantiation.*

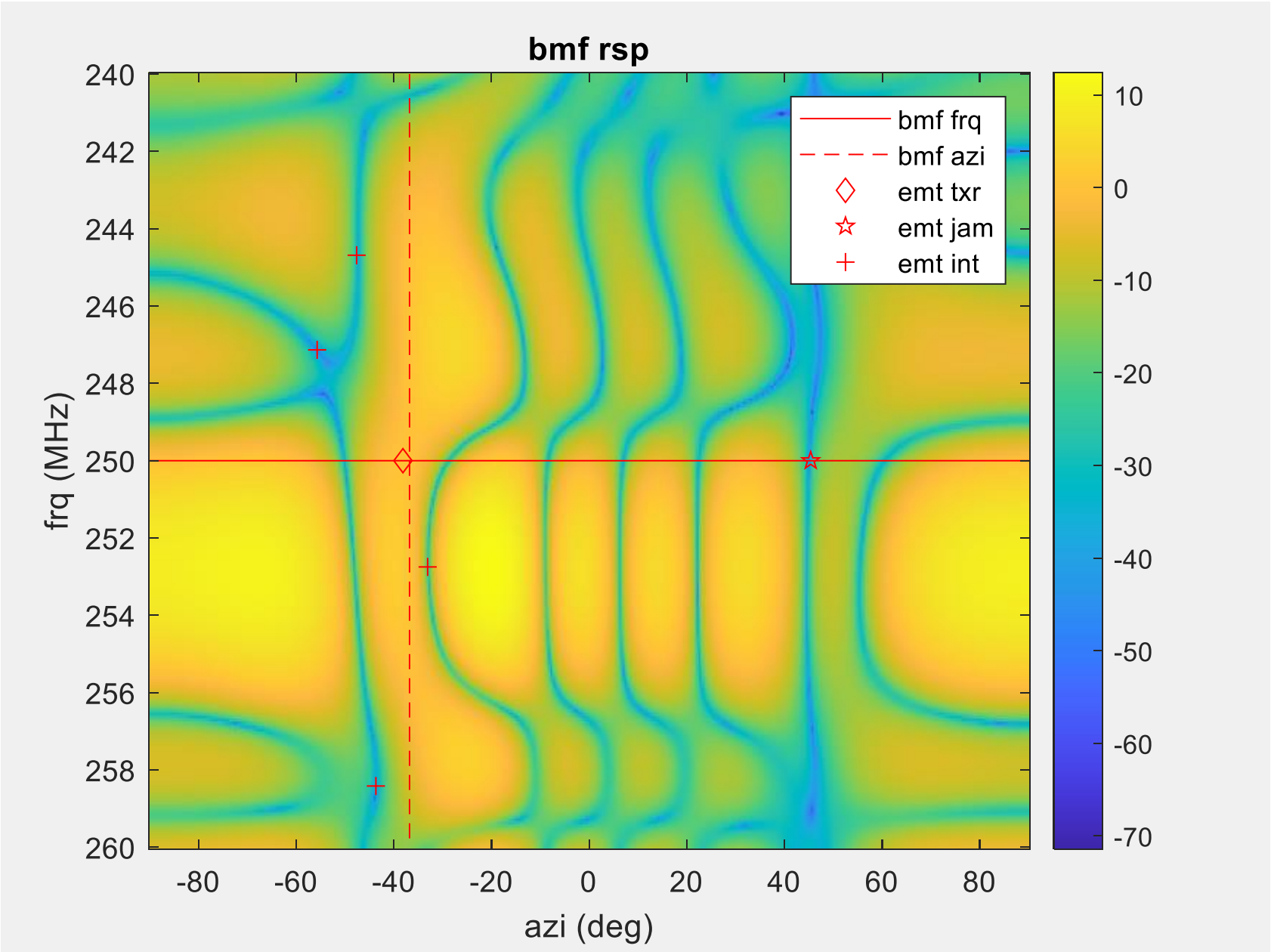

*Figure 4. Magnitude response (dB) of BMF-D; SNR = 19.0871 dB in this random instantiation.*

## V. Conclusion

This paper considers using the optimal group delay in LCMV beamformers. A procedure for the determination of the optimal group delay is presented. The simulation results, for the VHF communication and UHF radar scenarios, suggest that the proposed procedure has the potential to either improve SNR or reduce the group delay (at the expense of a slight SNR reduction), relative to the standard LCMV procedure, where a fixed delay/phase-response is assumed.

## References

[1] H. L.Van Trees, *Optimum Array Processing,* Wiley-Interscience, New York, 2002.

[2] O. L. Frost, "An algorithm for linearly constrained adaptive array processing", in *Proceedings of the IEEE*, vol. 60, no. 8, pp. 926-935, Aug. 1972. https://doi.org/10.1109/PROC.1972.8817

[3] R. Ebrahimi and S. R. Seydnejad. "Elimination of Pre-Steering Delays in Space-Time Broadband Beamforming Using Frequency Domain Constraints." *IEEE Communications Letters* vol. 17, no. 4, pp. 769–72, 2013. https://doi.org/10.1109/LCOMM.2013.022713.130090

[4] Zhang Ming Xiaojian Wang and Anxue Zhang. "An Efficient Broadband Adaptive Beamformer without Presteering Delays." *Sensors (Basel, Switzerland)* vol. 21, no. 4, p 1100, 2021. https://doi.org/10.3390/s21041100.

[5] Yong Zhao, Wei Liu, and R J Langley. "Adaptive Wideband Beamforming With Frequency Invariance Constraints" *IEEE Transactions on Antennas and Propagation*, vol. 59, no. 4, pp. 1175–84, 2011. https://doi.org/10.1109/TAP.2011.2110630

[6] C.-Y. Tseng, "Minimum Variance Beamforming with Phase-Independent Derivative Constraints", *IEEE Transactions on Antennas and Propagation*, vol. 40, no. 3, pp. 285–94, 1992. https://doi.org/10.1109/8.135471

[7] Julan Xie, Zishu He, Shenyu Rao, Ziyang Cheng and Huiyong Li. "Digital Adaptive Wideband Beamforming without Pre-Steering Delay Processing" *Digital Signal Processing,* vol. 132, 103813, 2023. https://doi.org/10.1016/j.dsp.2022.103813

[8] Yang Wang, Haiyun Xu, Bin Wang and Minglei Sun. "Optimal Array Phase Center Study for Frequency-Domain Constrained Space-Time Broadband Beamforming", *IEEE Access*, vol. 11, pp. 1–1, 2023. https://doi.org/10.1109/ACCESS.2022.3233645

[9] Shurui Zhang, Qiong Gu, Hui Sun, Weixing Sheng and Thia Kirubarajan, "Adaptive Broadband Frequency Invariant Beamforming Using Nulling-Broadening and Frequency Constraints", *Signal Processing,* vol. 195, 108461, 2022. https://doi.org/10.1016/j.sigpro.2022.108461

[10] L.C. Godara, and M.R.S Jahromi. "Convolution Constraints for Broadband Antenna Arrays", *IEEE Transactions on Antennas and Propagation,* vol. 55, no. 11, pp. 3146–54, 2007. https://doi.org/10.1109/TAP.2007.908823

[11] Yang Wang, Haiyun Xu, Bin Wang, and Minglei Sun. "Broadband Spatial Spectrum Estimation Based on Space-Time Minimum Variance Distortionless Response and Frequency Derivative Constraints", *IEEE Access* vol. 11, 1–1, 2023. https://doi.org/10.1109/ACCESS.2023.3258978

[12] H.L. Kennedy, "Recursive Optimal Filters for Smoothing and Prediction with Instantaneous Phase and Frequency Estimation Applications" *Circuits Syst Signal Process,* (to appear), 2025. https://doi.org/10.1007/s00034-025-03155-0

[13] I.W Selesnick and C.S Burrus. "Maximally Flat Low-Pass FIR Filters with Reduced Delay", *IEEE Transactions on Circuits and Systems. 2, Analog and Digital Signal Processing,* vol. 45, no. 1, pp. 53–68, 1998. https://doi.org/10.1109/82.659456

[14] O. Herrmann, "On the Approximation Problem in Nonrecursive Digital Filter Design", *IEEE Transactions on Circuit Theory,* vol. 18, no. 3, pp 411–13, 1971. https://doi.org/10.1109/TCT.1971.1083275

[15] Olivia Zacharia, and Vani Devi M. "Performance Analysis of OTFS Signal with Different Pulse Shapes for JCR Systems" in Proc. *IEEE 18th International Colloquium on Signal Processing & Applications (CSPA)*, pp. 24–29, 2022. https://doi.org/10.1109/CSPA55076.2022.9781902